\documentclass[%
reprint,
 amsmath,amssymb,
 aps,
prb,
]{revtex4-2}

\usepackage{graphicx}
\usepackage{booktabs} 
\usepackage[mathlines]{lineno}
\usepackage{dcolumn}
\usepackage{bm}
\usepackage{hyperref}
\usepackage{xcolor}

\begin{document}

\preprint{APS/123-QED}

\title{Studying temperature-induced decrease of spontaneous polarization in PbTiO$_3$ from vibrational amplitude}

\author{Zhikai Gao}
\affiliation{%
Institute for Applied Physics, University of Science and Technology Beijing, Beijing 100083, China}%
\author{Yuanji Xu}%
\affiliation{%
Institute for Applied Physics, University of Science and Technology Beijing, Beijing 100083, China}

\author{Fuyang Tian}%
 \email{fuyang@ustb.edu.cn}
\affiliation{%
Institute for Applied Physics and State Key Laboratory for Advanced Metals and Materials, University of Science and Technology Beijing, Beijing 100083, China
}%
\date{\today}

\begin{abstract}
\begin{description}
\item[Abstract]
PbTiO$_3$ is a typical ferroelectric polarization induced negative thermal expansion (NTE) material at a wide range of temperature. 
The elegant TiO$_6$ octahedron is used to study the polarization and NTE, whereas lacking its quantitative characterization. In this work, we propose a concise projection of vibrational amplitude to describe anisotropic distortion of atoms with temperature based on $ab$ $initio$ molecular dynamics simulations. We find that the huge negative mode Grüneisen coefficient is mainly from the different vibrational modes of the TiO$_6$ octahedral rotation at two boundary symmetry points in tetragonal phase along $c$ direction. The potential energy surface as a function of amplitude suggests the $c$-axis compression facilitates the TiO$_6$ octahedral rotation. 
We show that the vibrational mode of TiO$_6$ octahedral rotation around negative-expansion axis can promote spontaneous polarization reduction with increasing temperature, making transformation of PbTiO$_3$ from tetragonal to cubic phase occur, via the projected amplitude, potential energy surface, and forces of individual atoms.

\item[Keywords]
Polarization; Negative thermal expansion; Vibrational amplitude; PbTiO$_3$; $Ab$ $initio$ calculations
\end{description}
\end{abstract}

\maketitle


In perovskite ABO$_3$ and perovskite-like materials, different structural distortions occur due to the different atomic radii and valence states of A and B sites, resulting in interesting physical properties such as ferroelectricity, giant magnetoresistance, superconductivity, etc. In addition to the extensively studied Jahn-Teller distortion~\cite{jahn1937stability,bersuker2020jahn,varignon2019origins}, there exists a common structural distortion, namely the rotation/tilt of oxygen octahedra~\cite{glazer1972classification,adams2023classification}. Understanding the rotational form of oxygen octahedra at the atomic scale plays a key role in the correlation between structure and physical properties, such as spontaneous ferroelectric polarization influenced by oxygen octahedral rotation with temperature which is also a fundamental issue in ferroelectric materials.

PbTiO$_3$ is a classical perovskite ferroelectric material, exhibiting a tetragonal ferroelectric phase below its Curie temperature (763 K) and a cubic paraelectric phase above this temperature~\cite{shirane1956x}. Due to its large spontaneous polarization, high Curie temperature, and strong tetragonal distortion, PbTiO$_3$ has attracted extensive interest. 
Cohen~\cite{cohen1992origin} studied the microscopic origin of ferroelectricity from the perspective of electronic structure and demonstrate that hybridization between the 3$d$ states of Ti and the 2$p$ states of O is essential for ferroelectricity. 
The lattice dynamical instability plays a crucial role in the emergence of spontaneous polarization. In the classical soft-mode theory~\cite{cochran1959crystal,cochran1960crystal}, the spontaneous polarization originates from the freezing of a certain optical vibrational mode. The soft mode induces structure phase transition, which is classified as ferrodistortive (FD) or nonpolar antiferrodistortive (AFD) depending on the location of the unstable mode in the Brillouin zone. Ferroelectric (FE) distortion belongs to the FD, whereas typical AFD is associated with rotational instabilities of TiO$_6$ octahedron. Kornev $et$ $al$.~\cite{zhong1995competing,sai2000first,kornev2006phase,wojdel2013first} found that FE and AFD distortions could be coupled and generally compete with each other in perovskites. Gu $et$ $al$.~\cite{gu2018cooperative} further demonstrated that the competing relationship between FE and AFD distortions could also become cooperative in numerous perovskite oxides. 

However, the above study and similar works mainly focus on the emergence of polarization associated with the phase transition~\cite{king1994first,zhong1995first,waghmare1997ab,nishimatsu2010first,sharma2014first}, whereas the temperature evolution of spontaneous polarization within the stable tetragonal ferroelectric phase remains unclear. Via studying the pyroelectric effect in ferroelectric PbTiO$_3$, Ekulund $et$ $al$. demonstrated a dynamically stable phonon dispersion for tetragonal phase with different $c/a$ ratios at a wide range of temperature~\cite{eklund2024pyroelectricity}. It raises an interesting question on how the AFD soft-mode drives the spontaneous polarization. It is well worth exploring spontaneous polarization influenced by the finite-temperature TiO$_6$ octahedral rotational vibrations and their anharmonic coupling with the polar mode. Lin $et$ $al.$'s findings highlight the critical role of thermal vibrations in understanding ferroelectric phase transition mechanisms in perovskite oxides when the electron energy-loss spectroscopy (EELS) was used to investigate the thermal vibrations of oxygen atoms in PbTiO$_3$ and SrTiO$_3$ ~\cite{lin2024isotropic}. Thus it is crucial to study the electronic and structural distortion as well as related properties of ferroelectric phase at low temperature. For example, Chen $et$ $al$. found the negative thermal expansion (NTE) behavior could be strongly associated with the spontaneous polarization~\cite{chen2011role,pan2026strong,pan2025anion}.

The NTE of PbTiO$_3$ persists over a broad temperature range from room temperature to the Curie temperature~\cite{shirane1950x,chen2005thermal}.
The phonon-driven NTE mechanism of tetragonal PbTiO$_3$ was extensively studied via the quasi-harmonic approximation (QHA) and Grüneisen theory ~\cite{wang2013first,wang2014first,jiao2016first,ritz2018interplay}. Especially, Ritz $et$ $al$.~\cite{ritz2018interplay,Mittal2019,Ritz2019} calculated anisotropic coefficients of thermal expansion (CTEs) as a function of temperature along three lattice directions by means of the lattice parameter and elastic constants at room temperature, achieving excellent agreement with the experiments of PbTiO$_3$. The uniaxial Grüneisen coefficients and elastic compliance tensors jointly determine the anisotropic NTE behavior, which are both essentially derived from the phonon dispersion. The elastic constants can be extracted from acoustic branches~\cite{yang2025htem}, while the negative Grüneisen coefficients along $c$ ($a/b$) axis is from optical (acoustic) branches. Thus the vibrational modes of phonons are a key factor for understanding thermal expansion and its related spontaneous polarization with temperature. 
The negative Grüneisen coefficient indicates that the phonon frequency decreases upon lattice compression. However, the origin of the negative mode Grüneisen coefficient, particularly the relationship between specific phonon vibrational mode and spontaneous polarization, remains insufficiently understood.

Here we introduce the anisotropic atomic amplitude to study the mechanism underlying the reduction of spontaneous polarization within the stable tetragonal phase below the Curie temperature. We find that the microscopic origin of the negative Grüneisen coefficient under compressed $c$ axis is associated with the nonpolar TiO$_6$ octahedral vibrational mode at two key high symmetry points along the boundary of Brillouin zone. The potential energy surface of vibrational mode for TiO$_6$ octahedra rotation can unveil its anomalous dependence of strain. When transiting from tetragonal to cubic phase, the vibrational amplitude of TiO$_6$ octahedron increases, whereas its frequency decreases.  
From the self-consistent phonon (SCPH)~\cite{werthamer1970self,tadano2015self} calculations at finite temperature and $ab$ $initio$ molecular dynamics (AIMD) \cite{frenkel1997understanding} simulations, we extract the mean square amplitude (MSA) and anisotropic amplitude of individual atoms to characterize the anharmonic potential landscape of TiO$_6$ octahedral vibrations. The atomic forces and non-symmetric potential energy surface further shed light on the vibrational-mode-induced reduction of spontaneous polarization with temperature. 

The MSA of phonon vibration mode $(\bold{k},j)$ satisfies~\cite{maradudin1971theory}
\begin{flalign}\label{eq7}
&\ \left< Q\binom{\bold{k}}{j}Q^*\binom{\bold{k'}}{j'}\right>=\frac{\bold{Tr}Q\binom{\bold{k}}{j}Q^*\binom{\bold{k'}}{j'}e^{-\beta H}}{\bold{Tr}e^{-\beta H}} &
\notag\\
&=\Delta(\bold{k}-\bold{k'})\delta_{jj'}\frac{\hbar}{2\omega_j(\bold{k})}\coth{\frac{1}{2}\beta\hbar\omega_j(\bold{k})},
\end{flalign}
where the normal coordinate $Q\binom{\bold{k}}{j}$ is the amplitude of branch $j$ at wave vector $\bold{k}$,
$\beta=\frac{1}{k_B T}$, $k_B$ is Boltzmann constant. $\Delta(\bold{k}-\bold{k'})$ and $\delta_{jj'}$ represent the Kronecker delta functions for wave vectors and phonon branches, respectively, ensuring the orthogonality of phonon normal modes. The bracket $<>$ indicates an average over the ensemble. 

From Eq. \ref{eq7}, we can see that the MSA of mode $\binom{\bold{k}}{j}$ can be calculated at given frequency of mode $\bold{k}$. The phonon frequency is not explicitly temperature-dependent, whereas the phonon frequency with temperature can be substituted into Eq. \ref{eq7} to provide an approximate description of the temperature dependence of MSA.
Note that the MSA from Eq. \ref{eq7} characterizes the total vibrational amplitude of a phonon mode and cannot provide the anisotropic vibration information of single atom. Theoretically, the radial spatial distribution of vibrational amplitude from thermal vibration can be calculated with the Debye or Einstein model \cite{Lonsdale1948,Zhou2024}. Unfortunately it is slightly difficult to calculate accurately the Debye/Einstein temperature at finite temperature \cite{Zhou2024}. And the simulated vibrational amplitude (often named as thermal vibration induced atomic displacement) is isotropic and the average value of all atoms. Although molecular dynamics (MD) simulations can account for anharmonic effects at elevated temperatures~\cite{nemausat2015phonon,nemausat2017temperature}, they do not directly provide the anisotropic vibrational amplitudes of individual atoms. 

To this end, we propose a simple approach to calculating the anisotropic atomic amplitude. 
Based on finite-temperature lattice vibration of configurations simulated by AIMD in the NVT ensemble (N is the number of particles, V is volume, and T is temperature of a system), we can extract vibrational amplitudes of individual atoms with the anharmonicity of interatomic interactions induced by temperature. This amplitude could be projected onto the three Cartesian axes. 

We define the statistical average amplitude of the $i$-th atom:
\begin{align}\label{eq13}
\boldsymbol{A_i}=\sum_{m}^{M}\left | \Tilde{\boldsymbol{u}}_{i,m} \right | \cdot D_m,
\end{align}
where $\boldsymbol{A_i}$ is a displacement vector, it's projection along three Cartesian axes is statistical average amplitude of $i$-th atom, $\Tilde{\boldsymbol{u}}_{i,m}$ for the effective deviation in $m$-th interval, and $D_m$ for the density of the number of displacement. The detailed derivation of Eqns. \ref{eq7} and \ref{eq13} and all $ab$ $initio$ calculations details as well as the accuracy tests (Figs.S1-3) are shown in Supplementary Materials (SM).


\begin{figure*}[htbp]
    \centering
    \includegraphics[width=0.85\textwidth]{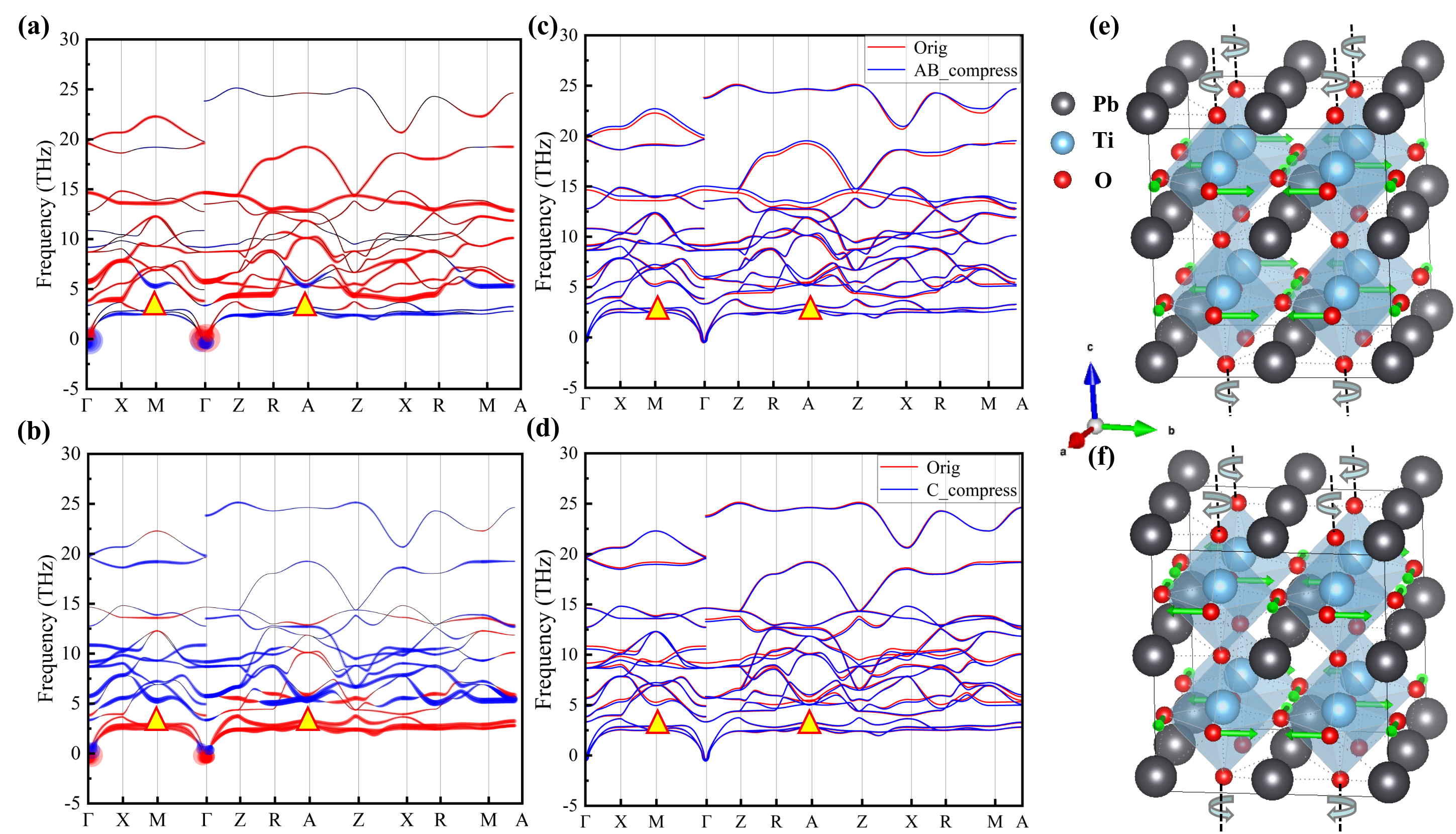}
    \caption{\label{Phonon} (a): Mode Grüneisen coefficients $\gamma^a_{q,j}=\gamma^b_{q,j}$ along highly symmetric path, band thickness proportional to the magnitude of $\gamma^a_{q,j}$ for each mode and color corresponding to sign of $\gamma^a_{q,j}$ (red positive, blue negative); (b): Mode Grüneisen coefficients $\gamma^c_{q,j}$ along highly symmetric path; (c): Comparison of phonon dispersion between the original and compressing $a$ and $b$ lattice parameter; (d): Comparison of phonon dispersion between the original and compressing $c$ lattice parameter; (e): Vibrational mode of oxygen octahedron at point M; (f): Vibrational mode of oxygen octahedron at point A.}
\end{figure*}

We set aside the controversial nature of the ferroelectric phase transition in PbTiO$_3$, particularly regarding whether it is first order or second order ~\cite{fang2015nature} within a narrow temperature interval around the Curie temperature. Away from the phase transition region, the $c/a$ ratio and volume of tetragonal structure evolve continuously with temperature, rather than exhibiting discontinuous changes. 

In order to quantitatively describe the uniaxial strain during the phase transition, we calculated the CTEs over the temperature range of 0-800 K using the QHA in combination with the Grüneisen theory and elastic calculations. The detailed results are shown in Fig.S4 and Fig. S5 of SM. The $a$ axis of PbTiO$_3$ exhibits PTE, with a CTE of 10.44$\times$10$^{-6}$ K$^{-1}$ at 300 K. At low temperature the freezing of optical phonons generates polarization, while the acoustic phonons are excited, resulting in PTE ~\cite{wang2014first}. 
In contrast, the $c$ axis exhibits NTE, with a calculated CTE of -24.03$\times$10$^{-6}$ K$^{-1}$ at 300 K. Our results are consistent with the experimental measurements~\cite{chen2008zero} and theoretical calculations~\cite{ritz2018interplay}, although minor discrepancies remain. It suggests that the accurate mode Grüneisen coefficients and elastic compliance coefficients at finite temperature are crucial for understanding phonon vibrational modes. 

The mode Grüneisen coefficient can be used to study the influence of uniaxial strain on vibrational modes. As shown in Fig. \ref{Phonon}(a) and (b), the mode Grüneisen coefficients along $a$ axis are negative for the low frequency acoustic phonon branches and for several optical branches around 5 THz, corresponding to the modes of negative mode Grüneisen coefficients at M and A points highlighted in Fig. \ref{Phonon}(a). Along $c$ axis, the negative mode Grüneisen coefficients are mainly distributed around 5 THz and the higher frequency optical phonon branches. Notably, the optical phonon branches above 5 THz account for the majority of the negative mode Grüneisen coefficients over the entire phonon spectrum. Among them, the low-frequency optical phonon branches near 5 THz at M and A points exhibit exceptionally large negative mode Grüneisen coefficients. For the $j$-th phonon branch with the wavevector $q$ along the crystallographic $a$ axis, $\gamma^a_{q,j}=-\frac{1}{\omega_{q,j}}\frac{\partial \ln{\omega_{q,j}}}{\partial \ln{\epsilon_a}}$ indicating that the mode Grüneisen coefficient characterizes the anharmonic coupling between lattice strain and phonon frequencies. A positive (negative) mode Grüneisen coefficient indicates that the phonon frequency decreases (increases) with increasing strain.

Figure \ref{Phonon}(c) and (d) compare the phonon dispersion of the equilibrium structure with those of the structures compressed along $a$ ($b$) and $c$ axes, respectively. The low frequency optical phonon branches (around 5 THz) at M and A points exhibit pronounced softening after compressing lattice parameters. This behavior is consistent with the negative mode Grüneisen coefficients shown in Fig. \ref{Phonon}(a) and (b). Figure \ref{Phonon}(e) illustrates the atomic displacement pattern of low frequency optical phonon branch (around 5 THz) at M point. In the vibrational mode, four oxygen atoms in the (001) planes surrounding the Ti atom undergo collective rotational vibrations about the Ti atom along $<$001$>$ direction. The oxygen atoms in the upper and lower (001) planes vibrate along the same direction, corresponding to a collective rotation motion of the TiO$_6$ octahedron. Therefore, this vibratio mode is referred to as the TiO$_6$ octahedral rotation mode. Figure \ref{Phonon}(f) shows the atom displacement pattern of the corresponding phonon branch at A point, which is similar to Fig. \ref{Phonon}(e), except that the TiO$_6$ octahedra in the upper and lower (001) planes rotate along opposite directions.

The TiO$_6$ octahedral rotational modes at M and A points have obvious negative mode Grüneisen coefficients. However, $\gamma^a_{q,j}$ only quantifies the relationship between strain and phonon frequency, which does not reveal the microscopic origin of the negative mode Grüneisen coefficients. Consequently, the physical mechanism responsible for the anomalous strain dependence of the TiO$_6$ octahedral rotational modes at M and A points remains unclear. 

\begin{figure}[htbp]
    \centering
    \includegraphics[width=0.85\linewidth]{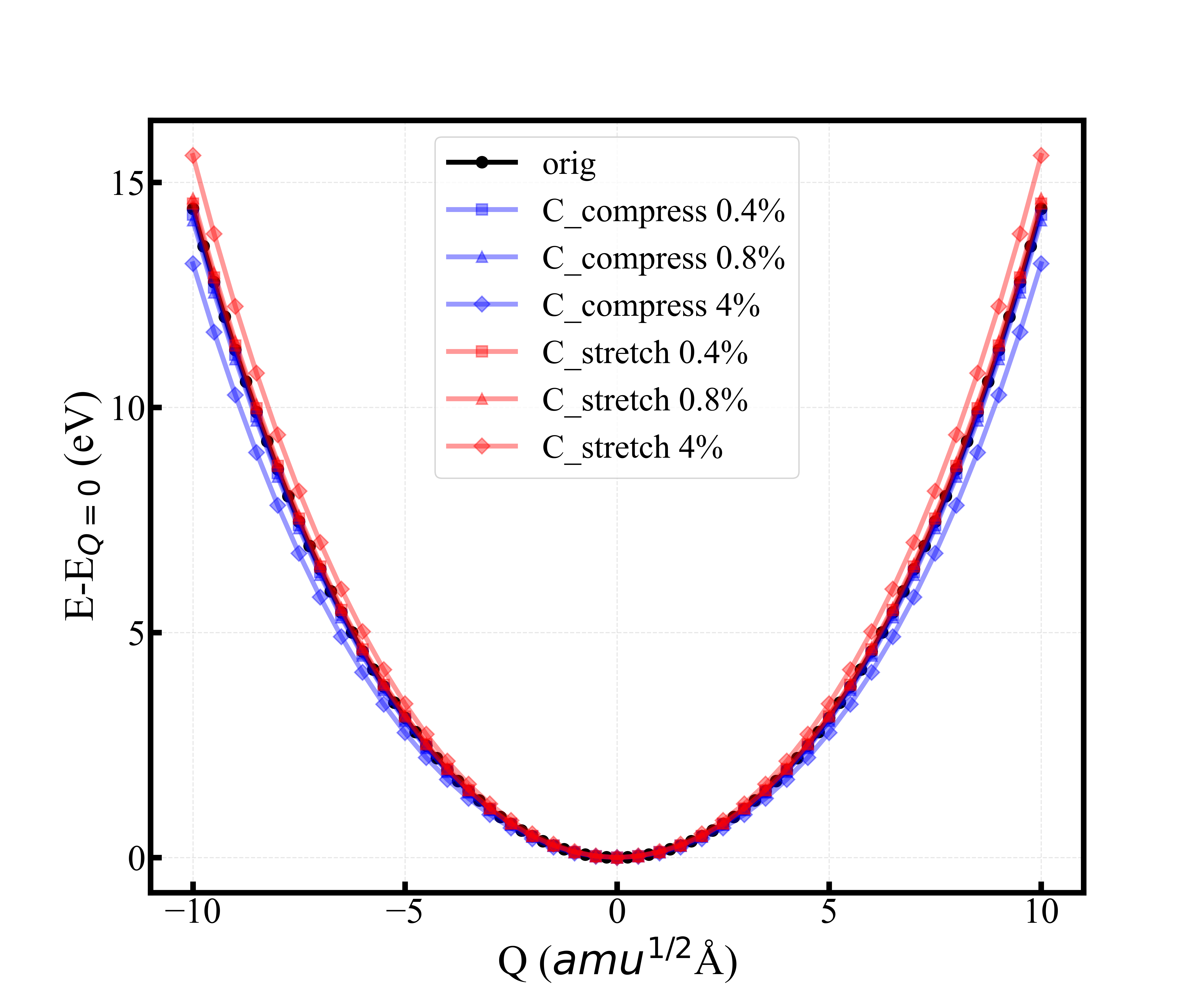}
    \caption{\label{Potential} Potential energy surface along M-point TiO$_6$ octahedral rotational mode as a function of amplitude (normal-mode coordinates) $Q$ under different compressive and tensile strains along $c$ axis. The definition of $Q$ is given in Eq. (2) of SM.}
\end{figure}

To this end, we employed the ModeMap software package~\cite{skelton2016anharmonicity} to calculate the potential energy surfaces of these two vibrational modes. For the TiO$_6$ octahedral rotational mode at M symmetry point, Fig. \ref{Potential} shows the single-well potential energy surfaces as a function of amplitude at different uniaxial strains. It indicates that the atoms corresponding to the vibrational mode sit on stable equilibrium positions of the tetragonal phase of PbTiO$_3$. The  potential energy surface of the corresponding vibrational mode at A point is shown in Fig. S6 of SM.
The compressing of $c$ axis requires less energy to achieve the same vibrational amplitude, whereas stretching of $c$ axis needs more energy at the same amplitude. It suggests that $c$ axis compression facilitates the rotation of TiO$_6$ octahedra, whereas the stretching $c$ axis suppresses the rotation of TiO$_6$ octahedra. A similar result is obtained for the TiO$_6$ octahedral rotational mode at A point, as shown in Fig. S6 of SM. This conclusion is consistent with the findings from Zhong $et$ $al$.~\cite{zhong1995competing}. 

The energy change is related to force: $F(x)=-\frac{\partial U(x)}{\partial x}$, where the negative sign indicates that the force is a restoring force. According to the harmonic oscillator model, the vibrational frequency is proportional to the square root of the force constant. Figure \ref{Potential} shows a decrease of restoring force when the $c$ axis is compressed, meaning the decrease of phonon frequency. It further shows the negative mode Grüneisen coefficients observed in the TiO$_6$ octahedral rotational modes at M and A points when the $c$ axis is compressed.


\begin{figure*}[htbp]
    \centering
    \includegraphics[width=0.95\textwidth]{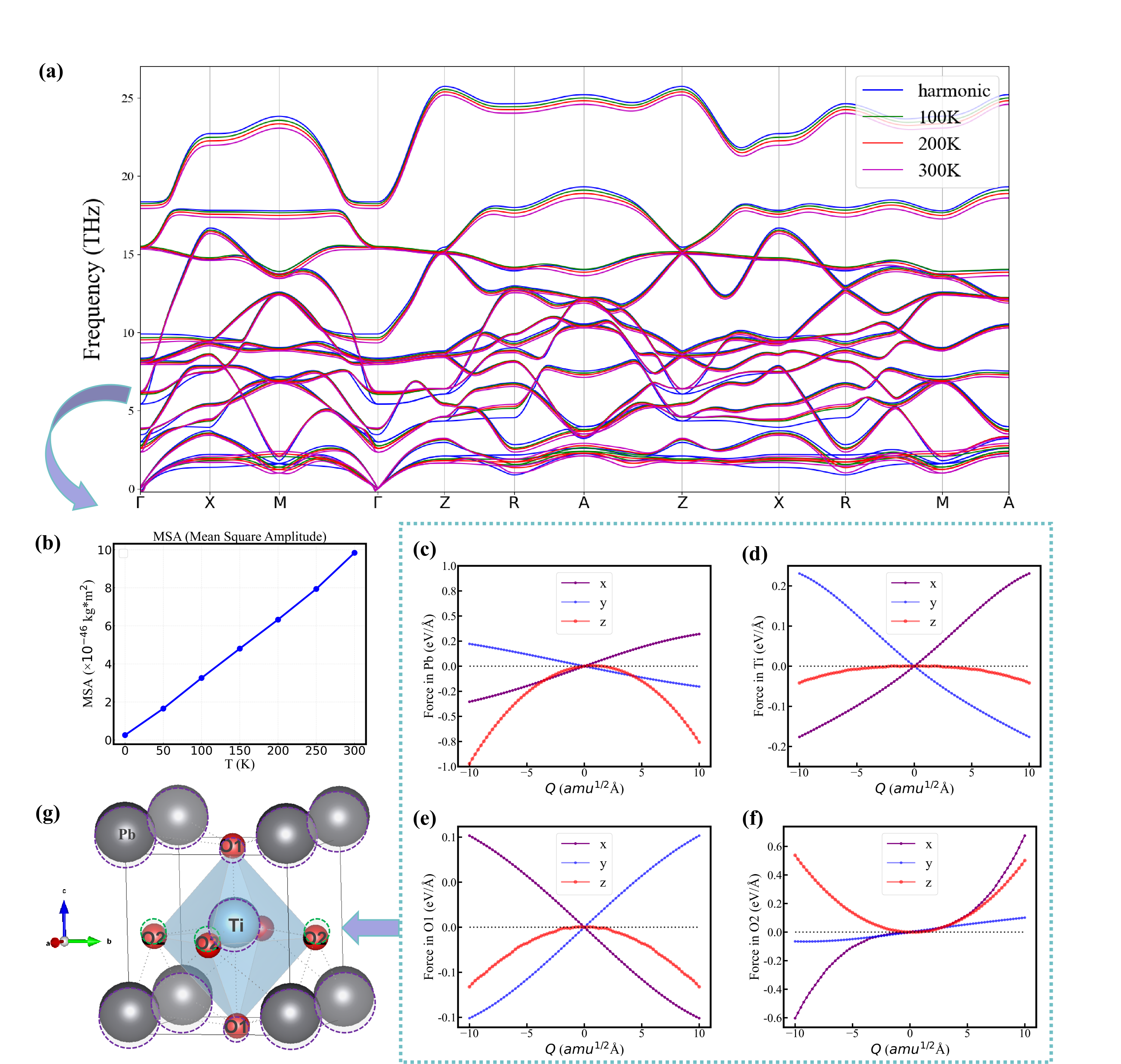}
    \caption{\label{Q_Force} (a): Finite temperature phonon dispersion compared with the harmonic phonon dispersion. (b): Temperature (T) dependence of the mean square amplitude (MSA). (c)-(f): Influence of M-point TiO$_6$ octahedral rotational mode on atomic forces, showing the forces acting on Pb, Ti, O1 and O2 atoms as a function of the normal-mode amplitude $Q$. The corresponding atoms are marked in Fig. \ref{Q_Force}(g). The definition of $Q$ is given in Eq. (2) of SM. (g): The structure of tetragonal PbTiO$_3$. The schematic dash circles suggest the influence of M-point TiO$_6$ octahedral vibration on the forces generated by each atom, leading to potential atomic movement processes.}
\end{figure*}

\begin{figure*}[htbp]
    \centering
    \includegraphics[width=\textwidth]{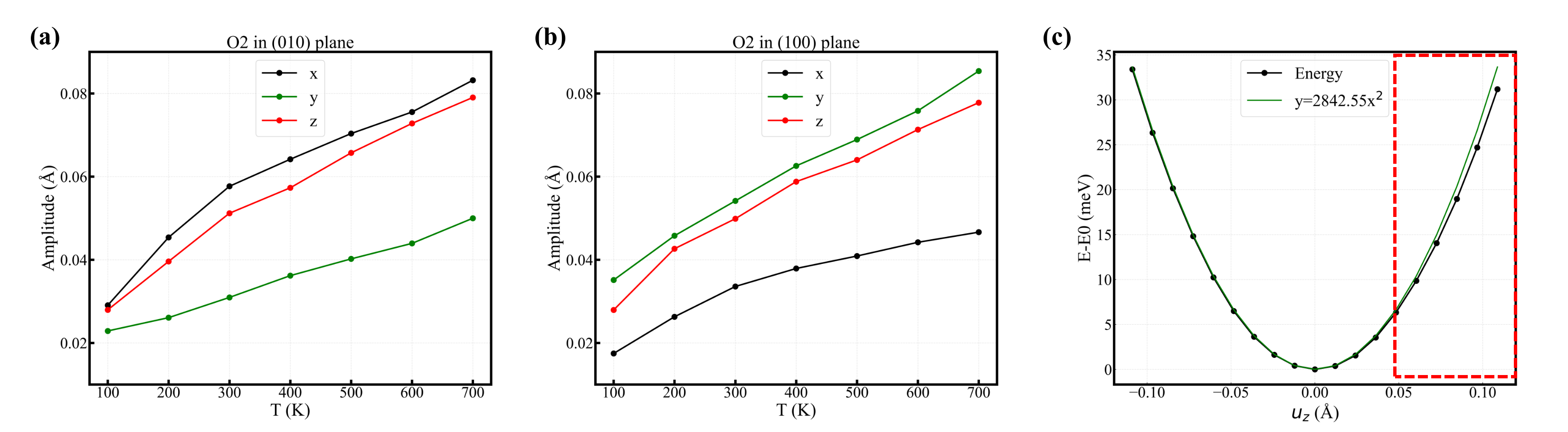}
    \caption{\label{Amplitude} (a)-(b): Temperature dependence of the O2 atoms in (010) and (100) planes (marked in Fig. \ref{Q_Force}(g)) vibrational amplitude along three directions. (c): Energy difference of the O2 atoms displaced along the positive and negative $z$ direction. Here $u$ denotes the amplitude (displacement) of O2 atom away from the equilibrium position. The green line represents the energy differences at $u_z < 0$ fitting to a quadratic function.}
\end{figure*}

During the transition from tetragonal phase to cubic phase, the FE soft mode associated with FE distortion occurs at the center of the Brillouin zone ($\Gamma$) and can directly modify the magnitude of polarization. In contrast, the AFD soft mode occurs at the Brillouin zone boundary and does not directly change the polarization. There is a coupling between FE and AFD modes~\cite{zhong1995competing,gu2018cooperative}. The TiO$_6$ octahedral rotational modes play an important role in understanding the decrease of polarization in PbTiO$_3$ during the phase transition. 

We have shown that the $c$-axis contraction could increase the vibrational amplitude of TiO$_6$ octahedron, while its vibrational frequency decreases. 
Different from discussion based on cubic phase of PbTiO$_3$, we investigate the influence of nonpolar TiO$_6$ octahedral rotational modes on polarization in tetragonal phase. From the temperature-dependent phonon dispersion shown in Fig. \ref{Q_Force}(a), these nonpolar vibrational modes of TiO$_6$ octahedron at M and A points do not exhibit imaginary frequency at finite temperature. We calculated the MSA of M-point vibrational mode by combining the temperature dependent phonon frequency and Eq. \ref{eq7}. As shown in Fig. \ref{Q_Force}(b),
the MSA increases with temperature, indicating that the vibrational amplitude of TiO$_6$ octahedron enlarges as the temperature increases. 

To investigate the influence of vibrational modes on polarization, we constructed a 3$\times$3$\times$3 supercell to displace the atoms along the TiO$_6$ octahedral rotational eigenvector. 
Figures \ref{Q_Force}(c)-(f) show the derived atomic forces as a function of vibrational amplitude on Pb, Ti, O1 and O2 atoms (see Fig. \ref{Q_Force}(g), these atoms are displaced along eigenvector of TiO$_6$ octahedral rotational mode at the M point).
A positive or negative amplitude Q represents two opposite directions of atomic displacement along the eigenvector of TiO$_6$ octahedral rotational mode. All atoms experience forces with the same magnitude but opposite directions along $x$ and $y$ directions when the vibrational amplitude Q has the same magnitude but opposite signs. It means that when the atom vibrates near equilibrium position in the vibrational mode of M point, The force projections of these atoms into $x$ and $y$ directions have a opposite direction, like a restoring force. Finally, the force along $x$ and $y$ directions cancel each other out and can't make atoms move away the equilibrium site. However, an anomalous behavior appears when the atomic forces are projected in $z$ direction. In this vibrational mode, the force projections on $z$ axis have the same direction, whenever atoms displacement around the equilibrium position along the positive or negative direction. The atomic forces acting on Pb, Ti, and O1 atoms are along negative $z$ direction, whereas the force acting on O2 atoms is along the positive direction of $z$ axis. The force projection along $z$ axis does not change with the sign of the amplitude and therefore does not behave as a conventional restoring force. The excitation of M-point vibration mode drives Pb, Ti, and O1 atoms toward the negative direction of $z$ axis and the O2 atoms toward the positive direction, resulting in atomic displacements away from their equilibrium positions, as shown by the shaded region in Fig. \ref{Q_Force}(g). Figure S7 in SM shows the derived atomic forces on Pb, Ti, O1 and O2 atoms along eigenvector of TiO$_6$ octahedral rotational mode at the A point.

These results are consistent with the structural evolution of PbTiO$_3$ during the phase transition. From the tetragonal to the cubic phase, O2 atoms shift from their off-center positions toward the centrosymmetric sites in (010) plane, while Ti atoms also move toward their centrosymmetric sites. Therefore, the TiO$_6$ octahedral rotational mode promotes the transformation of PbTiO$_3$ from tetragonal phase to cubic phase, that is, it promotes the reduction of spontaneous polarization.

For the mechanism underlying the decrease of spontaneous polarization with increasing temperature in PbTiO$_3$, we used AIMD simulations to calculate the temperature dependence of vibration amplitudes of O2 atoms along three crystalline directions, as shown in Fig.~\ref{Amplitude}(a) and (b). Results show that with increasing temperature, the vibrational amplitude of O2 atoms in (010) plane along the $x$ and $z$ axes increases, while the amplitude along the $y$ direction exhibits only a slight increase. The vibrational amplitude of O2 atoms in (100) plane along the $y$ and $z$ axes increases but $x$ direction increase slowly. It indicates that the amplitude of TiO$_6$ octahedron is facilitated with temperature increase. Based on the atomic force distribution shown in Fig.~\ref{Q_Force}(c)-(f), the increase amplitude of the TiO$_6$ octahedron enhances the magnitude of force acting on O2 and other atoms, the direction of forces is along the opposite $z$ axis. The induced relative displacement of O2 atoms with respect to other atoms along $z$ direction, thereby reducing the polarization.

To provide a more intuitive description of the origin for the relative movement of the O2 atom with other atoms along the positive $z$ axis, we calculated the potential energy curve of the O2 atom displaced from the equilibrium site along both positive and negative $z$ axis direction. We constructed a 3$\times$3$\times$3 supercell and displaced only one O2 atom along the positive and negative directions of $z$ axis. Figure \ref{Amplitude}(c) indicates that the potential energy curve for the O2 atom displaced along the $z$ axis direction is not symmetric. In contrast, the potential energy curves for displacements along $x$ and $y$ directions exhibit symmetric behavior (Fig. S8 of SM). According to $F(x)=-\frac{\partial U(x)}{\partial x}$, the slope of potential energy $U(x)$ curve with respect to displacement $x$ reflects the magnitude of restoring force, the minus sign is direction of force. Figure ~\ref{Amplitude}(c) suggests that the restoring force acting on O2 when displacing along the negative $z$ direction is larger than that along the positive direction of $z$ axis. It indicates that the magnitude of amplitude of O2 along the positive and negative directions of $z$ axis are different. Specifically, the amplitude along the positive $z$ direction is larger than that along the negative $z$ direction. Figure \ref{Amplitude}(a) and (b) shows that the vibrational amplitude of O2 along $z$ directions increase rapidly with temperature. The nonsymmetric atomic displacement along $z$ direction is enhanced with increasing temperature, inducing the equilibrium position (center of vibration) of O2 atom to shift toward the positive $z$ direction and resulting in a reduction of the spontaneous polarization.

In summary we have studied the phonon-mediated origin underlying the temperature-induced polarization decrease in tetragonal PbTiO$_3$ by our approach to projection of vibrational amplitudes, based on the combination of Grüneisen theory and $ab$ $initio$ molecular dynamics simulations. 
The microscopic mechanism of the coupling between TiO$_6$ octahedral vibrations, lattice strain, and polarization evolution is revealed in tetragonal PbTiO$_3$. The TiO$_6$ octahedral rotational modes show strong sensitivity to lattice strain, and their frequencies decrease under compression, resulting in negative mode Grüneisen coefficients. The potential energy surfaces as a function of amplitude at different strains further clarify the origin of these negative mode Grüneisen coefficients. The mean square amplitudes of the vibrational modes can be extracted from temperature dependent phonon dispersion data. Our anisotropic vibrational amplitudes from AIMD simulations quantify the magnitude of anharmonic effect increases with temperature. The temperature excites the increase of vibrational amplitudes, inducing atomic displacements along $z$ direction, as confirmed by atomic force analysis. The nonsymmetric potential energy surface further shows that increasing O2 displacement amplitude along $z$ direction, contributing to the reduction of spontaneous polarization in tetragonal PbTiO$_3$.

Our proposed anisotropic amplitude could be a new quantitative parameter to characterize the microscopic origin of the interplay between anharmonic lattice dynamics and ferroelectric polarization. 
Our findings provide a strategy for in-depth understanding of perovskite-like materials with coupled thermal expansion and ferroelectric properties.

\textbf{Acknowledgement:} This work was supported by the Science Challenge Project (No. TZ2025009), the National Natural Science Foundation of China (Grants No. 52371174 and No. 12204033), and the State Key Laboratory of Advanced Metals and Materials (No. 2025-Z22).\\



\bibliography{apssamp}

\end{document}


\title{Supplementary material for “Studying temperature-induced decrease of spontaneous polarization in PbTiO$_3$ from vibrational amplitude”}

\author{Zhikai Gao}
\affiliation{%
Institute for Applied Physics, University of Science and Technology Beijing, Beijing 100083, China}%
\author{Yuanji Xu}%
\affiliation{%
Institute for Applied Physics, University of Science and Technology Beijing, Beijing 100083, China}

\author{Fuyang Tian}%
 \email{fuyang@ustb.edu.cn}
\affiliation{%
Institute for Applied Physics and State Key Laboratory for Advanced Metals and Materials, University of Science and Technology Beijing, Beijing 100083, China
}%
\date{\today}

\maketitle
In this supplementary material, we in details describe the derivation on Mean square amplitude, the projection of amplitude along three axes and thermal expansion coefficient based on quasi harmonic approximation，$ab$ $initio$ calculations, temperature-dependent phonon dispersion and accuracy test such as testing of temperature-dependent phonon dispersion in the section of Methodology and calculated details \ref{method}. 

In section of Results \ref{result}, we show the quasi harmonic approximation calculation of NTE and A point vibrational mode of TiO$_6$ octahedron and amplitude projection.

\section{\label{method}Methodology and Calculated Details}
\subsection{Mean square amplitude}
The Hamiltonian for a crystal can be defined as
\begin{align}\label{eq4}
H=\Phi_0
&+\frac{1}{2}\sum_{l\kappa\alpha}M_{\kappa}\dot{u}_{\alpha}^{2}\binom{l}{\kappa}
\notag\\
&+\frac{1}{2}\sum_{\substack{l\kappa\alpha\\ l'\kappa'\beta}}\Phi_{\alpha\beta}\binom{ll'}{\kappa\kappa'}u_{\alpha}\binom{l}{\kappa}u_{\beta}\binom{l'}{\kappa'},
\end{align}
where the $\Phi_0$ is the static equilibrium potential energy, $M_{\kappa}$ for the mass of $\kappa$th kind of atom, $u_{\alpha}\binom{l}{\kappa}$ for the displacement of $\kappa$th atom in the $l$th unit cell from its equilibrium position along the $\alpha$ direction. And $\Phi_{\alpha\beta}\binom{ll'}{\kappa\kappa'}=\left(\frac{\partial^2\Phi}{\partial u_{\alpha}\binom{l}{\kappa}\partial u_{\beta}\binom{l'}{\kappa'}} \right)_0$ is the second derivative of energy with respect to displacement at equilibrium configuration. The solution in the form of a plane waves with the introduction of normal coordinates is:
\begin{align}\label{eq5}
u_{\alpha}\binom{l}{\kappa}=\frac{1}{\sqrt{NM_{\kappa}}}\sum_{\bold{k},j}e_{\alpha}\left( \kappa|\substack{\bold{k}\\j}\right)Q\binom{\bold{k}}{j}e^{2\pi i\bold{k}\cdot\bold{x}(l)},
\end{align}
where $e_{\alpha}\left(\kappa|\substack{\bold{k}\\j}\right)$ is the $\alpha$th Cartesian component of the unit vector for atom $\kappa$ in branch $j$ at wave vector $\bold{k}$. $Q\binom{\bold{k}}{j}$ is normal-mode coordinates (amplitude), the unit is $amu^{\frac{1}{2}}$\AA. It is the amplitude of branch $j$ at wave vector $\bold{k}$. Using the normal coordinate transformation given by Eq. \ref{eq5}, we have
\begin{align}\label{eq6}
\left< u_{\alpha}^{2}\binom{l}{\kappa}\right>
&=\frac{1}{NM_{\kappa}}\sum_{\bold{k}j}\sum_{\bold{k'}j'}\left< Q\binom{\bold{k}}{j}Q^*\binom{\bold{k'}}{j'}\right>
\notag\\
&e_{\alpha}\left(\kappa|\substack{\bold{k}\\j}\right)e_{\alpha}^*\left(\kappa|\substack{\bold{k'}\\j'}\right)\exp{\left[2\pi i(\bold{k}-\bold{k'})\cdot\bold{x}(l)\right]},
\end{align}
where the brackets $<>$ indicate an average over the canonical ensemble, of which N is the number of particle, V is volume, and T is temperature of a system. The mean square amplitude (MSA) of phonon mode $(\bold{k},j)$ satisfies\cite{maradudin1971theory}:
\begin{flalign}\label{eq7}
\left< Q\binom{\bold{k}}{j}Q^*\binom{\bold{k'}}{j'}\right>
&=\frac{\bold{Tr}Q\binom{\bold{k}}{j}Q^*\binom{\bold{k'}}{j'}e^{-\beta H}}{\bold{Tr}e^{-\beta H}} &
\notag\\
&=Z^{-1}\sum_n \left< n| Q\binom{\bold{k}}{j}Q^*\binom{\bold{k'}}{j'}|n \right>e^{-\beta E_n}
\notag\\
&=z_{\bold{k}j}^{-1}\Delta(\bold{k}-\bold{k'})\delta_{jj'}\sum_{n_j(\bold{k})=0}^{\infty}\frac{\hbar}{\omega_j(\bold{k})}\left( n_j(\bold{k})+\frac{1}{2}\right)\exp[-(n_j(\bold{k})+\frac{1}{2})\beta\hbar\omega_j(\bold{k})]
\notag\\
&=\Delta(\bold{k}-\bold{k'})\delta_{jj'}\frac{\hbar}{2\omega_j(\bold{k})}\coth{\frac{1}{2}\beta\hbar\omega_j(\bold{k})},
\end{flalign}
$\Delta(\bold{k}-\bold{k'})$ and $\delta_{jj'}$ represent the Kronecker delta functions for wave vectors and phonon branches, respectively, ensuring the orthogonality of phonon normal modes. Where $z_{\bold{k}j}$ is the partition function for the mode $\binom{\bold{k}}{j}$. $\beta=\frac{1}{k_B T}$, $k_B$ is Boltzmann constant.

From Eq. \ref{eq7}, we can see that the MSA of mode $\binom{\bold{k}}{j}$ can be calculated at given frequency of mode $\binom{\bold{k}}{j}$. The phonon frequency in Eq. \ref{eq7} is not explicitly temperature-dependent, whereas the phonon frequency with temperature can be substituted into Eq. \ref{eq7} to provide an approximate description of the temperature dependence of MSA.

\subsection{\label{Sec_amplitude}Projection of amplitude along three axes}
The MSA calculated from Eq. \ref{eq7} only characterizes the total vibrational amplitude of a phonon mode and cannot provide the anisotropic vibration information of single atom. Theoretically, the radial spatial distribution of vibrational amplitude from thermal vibration was calculated with Debye or Einstein model. Unfortunately it is slightly difficult to get accurately the Debye/Einstein temperature at finite temperature. And the simulated vibrational amplitude of atom is isotropic. Although the MD simulations can account for anharmonic effects at elevated temperatures \cite{nemausat2015phonon,nemausat2017temperature}, they do not directly provide the anisotropic vibrational amplitudes of individual atoms. 

We propose a simple approach to calculating the anisotropic atomic amplitude. Based on finite temperature lattice vibration of configurations simulated by AIMD at the NVT ensemble, we could extract atomic vibrational amplitudes with the anharmonicity of interatomic interactions induced temperature. This amplitude could be projected onto the three Cartesian axes.

For AIMD simulation at a certain temperature, we collect equilibrium configurations at different time steps. In a total of $N$-step AIMD simulations, the equilibrium position of the $i$-th atom $\bold{R}_i$ is 
\begin{equation}\label{eq8}
\boldsymbol{R}_i=\frac{1}{N}\sum_{n=1}^{N}\boldsymbol{r}_{i,n},
\end{equation}
where $\boldsymbol{r}_{i,n}$ denotes the position vector of the $i$-th atom at the $n$-th step of the molecular dynamics trajectory. The displacement of atoms away from the equilibrium position at each time step:
\begin{align}\label{eq9}
\boldsymbol{u}_{i,n}=\boldsymbol{r}_{i,n}-\boldsymbol{R}_i,
\end{align}
Let us sort the displacement $u_{i,n}$ and divide into M equally intervals. $\Delta \boldsymbol{u}$ is defined as 
\begin{align}\label{eq10}
\boldsymbol{\Delta u}=\frac{\boldsymbol{u}_{max}-\boldsymbol{u}_{min}}{M},
\end{align}
where $u_{max}$ and $u_{min}$ represent the maximum and minimum of displacement. Then we calculate the density of the number of displacement contained in the $m$-th interval $\boldsymbol{u}_{i,n}$-$\boldsymbol{u}_{i,n}+\Delta \boldsymbol{u}$:
\begin{align}\label{eq11}
D_m=\frac{N_m}{N},
\end{align}
where the $N_m$ is the number of the deviation in m-th interval. We define the effective deviation in m-th interval
\begin{align}\label{eq12}
\Tilde{\boldsymbol{u}}_{i,m}=\frac{1}{N_m}{\sum_{n_m=1}^{N_m}\boldsymbol{u}_{i,n}},
\end{align}
Finally, we define the statistical average amplitude of the $i$-th atom:
\begin{align}\label{eq13}
\boldsymbol{A_i}=\sum_{m}^{M}\left | \Tilde{\boldsymbol{u}}_{i,m} \right | \cdot D_m,
\end{align}
where $\boldsymbol{A_i}$ is a displacement vector, and it's projection along three Cartesian axes is statistical average amplitude of $i$-th atom. Introducing the effective deviation $\Tilde{\boldsymbol{u}}_{i,m}$ is for intuitively show the impact of the numerical distribution (or numerical density $D_m$) of the deviation $u_{i,n}$ on the total amplitude $A_i$, although the amplitude is mathematically equivalent to $\boldsymbol{A_i}=\frac{1}{N}\sum_{n=1}^{N}\left | \boldsymbol{u}_{i,n} \right |$ according to Eq. \ref{eq11}, \ref{eq12} and \ref{eq13}.

To calculate the amplitude projections, a 3$\times$3$\times$3 supercell was constructed. Equilibrium configurations at temperatures ranging from 100 to 700 K (with a temperature interval of 100 K) were obtained from AIMD simulations in the NVT ensemble. The detailed computational parameters are described in Sec. \ref{ab initio}. The vibration amplitudes along the three crystallographic axes were subsequently calculated from these configurations using the amplitude model introduced above.

\subsection{Thermal expansion coefficient based on quasi harmonic approximation}

Within the framework of the quasi harmonic approximation Grüneisen theory~\cite{gruneisen1926zustand,barron1980thermal,schelling2003thermal}, the axial CTE of a material is expressed as
\begin{align}\label{Eq_CTE}
    \alpha_l=\frac{C_\mathrm{V}}{V_0}\sum_kS_{l,k}\overline{\gamma_k},\, l,k=1,2,3,
\end{align}
where $C_\mathrm{V}$ is the constant volume specific heat, $V_0$ is the equilibrium volume, $S_{l,k}$ is the compliance tensor (i.e., the inverse of the elastic constants), $l, k$ is different directions. $\overline{\gamma_k}$ is the overall uniaxial Grüneisen parameter
\begin{align}\label{Eq_OverallGruneisen}
    \overline{\gamma_k}=\frac{\sum_{q,j}C_{q,j}\gamma^k_{q,j}}{\sum_{q,j}C_{q,j}},\, k=1,2,3,
\end{align}
where $C_{q,j}$ is the mode specific heat capacity of the $j$-th phonon mode with the wave vector $q$. And $\gamma^k_{q,j},\, k=1$ represents the Grüneisen parameter of the $j$-th phonon mode with the wavevector $q$ along the crystallographic $a$ axis. It represents the relation between frequency $\omega_{q,j}$ and strain $\epsilon_a$ along a axis (change at equilibrium position)
\begin{align}\label{Eq_modeGruneisen}
    \gamma^a_{q,j}=-\frac{1}{\omega_{q,j}}\frac{\partial \ln{\omega_{q,j}}}{\partial \ln{\epsilon_a}}.
\end{align}

From Eq. \ref{Eq_CTE}, the axial CTE requires to calculate the specific heat capacity, elastic constants (flexibility coefficients), and Grüneisen parameter of a material. The specific heat capacity can be directly calculated from the phonon spectrum data. The elastic constants were calculated based on the stress-strain relation implemented into HTEM software package~\cite{yang2025htem}. The Grüneisen parameter was calculated separately from the phonon spectrum data of the equilibrium unit cell and the lattice parameters of compressed and expanded unit cells, and then obtained using Eq. \ref{Eq_modeGruneisen}. We used the Phonopy software package~\cite{togo2023first,togo2023implementation} and combined it with our data post-processing code~\cite{gao2025high} to calculate the axial CTE. The ratio of expanding and contracting lattice parameters when calculating the Grüneisen parameter in the calculation mode is 0.5$\%$, and q-mesh is 30$\times$30$\times$30. We note that the negative thermal expansion of PbTiO$_3$ was excellently calculated via the above key input parameter based on lattice constants at room temperature~\cite{ritz2018interplay}.

\subsection{\label{ab initio}$Ab$ $initio$ calculations}

$Ab$ $initio$ calculations based on density functional theory are performed using the Vienna Ab initio Simulation Package (VASP) software~\cite{kresse1996efficiency}. During structural optimization, we validated generalized-gradient approximation (GGA) and local density approximation (LDA)~\cite{perdew1981self,perdew1996generalized} exchange correlation functionals. As show in Table S1, the optimized results of LDA functional were close to the experiments. Where the valence electrons of each atom is Ti 3s$^{2}$3p$^{6}$4s$^{2}$3d$^{2}$, Pb 5d$^{10}$6s$^{2}$6p$^{2}$ and O 2s$^{2}$2p$^{2}$. The interaction between core electrons and valence electrons is schemed through the projector augmented wave (PAW) method~\cite{blochl1994projector,kresse1999ultrasoft}. As show in Fig. \ref{Ecut_Kpot}(a) of the Supplementary Material (SM), we tested the kinetic energy cutoff for the plane-wave expansion and it was set to 600 eV. The threshold for the total energy convergence is 10$^{-8}$ eV and the convergence standard of force is 10$^{-3}$ eV/{\AA}. According to Fig. \ref{Ecut_Kpot}(b), the Brillouin zone integration is sampled by 9$\times$9$\times$8 Monkhorst-Pack k points.

\begin{table}[htbp]
\caption{\label{Table S1}: Lattice parameters of PbTiO$_3$ using different functionals, in unit of \AA.}
\begin{tabular}{@{}cccc@{}}
\toprule
Methods                   & a    & c         & c/a       \\ \midrule
PBE                & 3.8421 & 4.7456 & 1.2352 \\
LDA                & 3.8643 & 4.0241 & 1.0355 \\
Exp.~\cite{rossetti1998phase}  & 3.8997 & 4.1529 & 1.0649 \\ \bottomrule
\end{tabular}
\end{table}

\begin{table}[htbp]
\caption{\label{Table S2}: The coordinates of high symmetry points.}
\begin{tabular}{@{}ll@{}}
\toprule
Label & Coordinates   \\ \midrule
$\Gamma$    & (0,0,0)       \\
X     & (0,0.5,0)     \\
M     & (0.5,0.5,0)   \\
Z     & (0,0,0.5)     \\
R     & (0,0.5,0.5)   \\
A     & (0.5,0.5,0.5) \\ \bottomrule
\end{tabular}
\end{table}

\begin{table}[htbp]
\caption{\label{Tab_felxiblity}: Flexibility coefficient of tetragonal PbTiO$_3$.}
\begin{tabular}{cccc}
\toprule
S11(10$^{-12}$Pa$^{-1}$) & S12    & S13    & S33    \\ \midrule
4.324          & -0.778 & -2.720 & 11.642 \\ \bottomrule
\end{tabular}
\end{table}

\begin{figure}[H]
    \centering
    \includegraphics[scale=0.6]{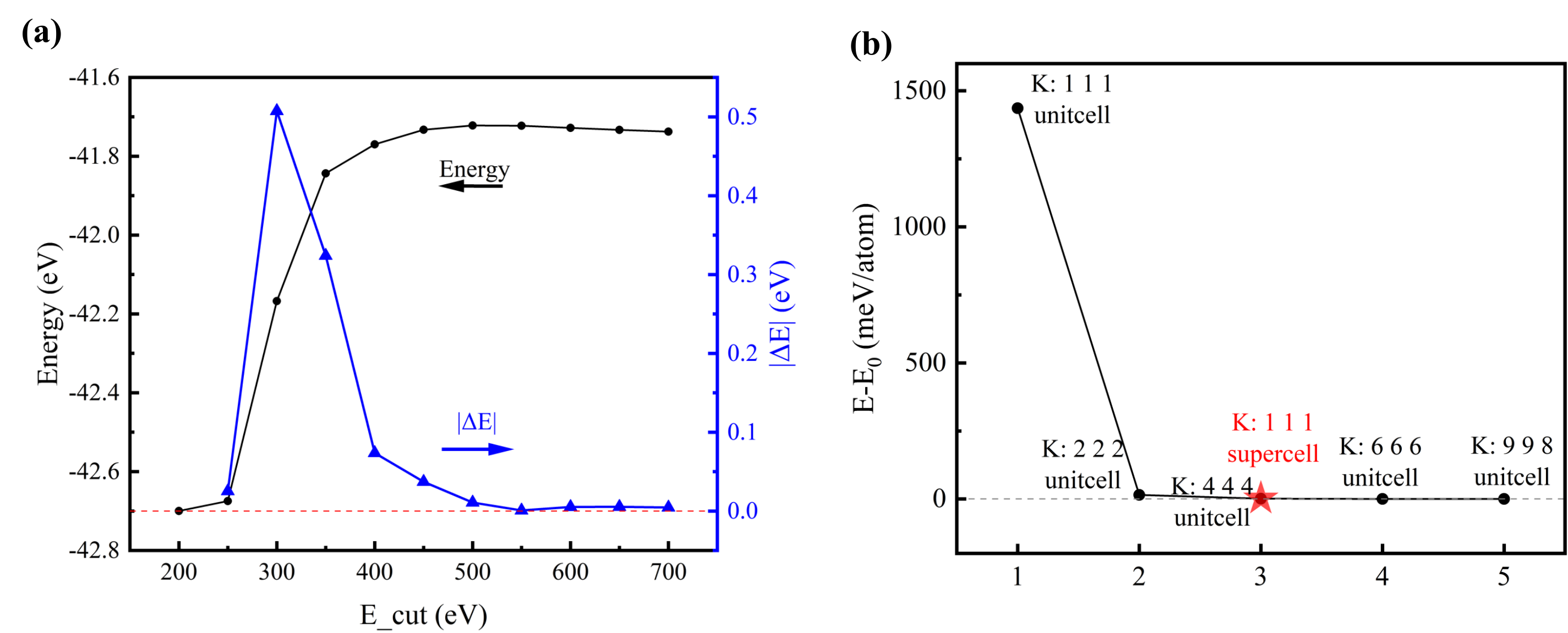}
    \caption{\label{Ecut_Kpot} (a): Cutoff energy convergence test. (b): K point convergence test. The black line represents the energy differences relative to the reference energy calculated using different k-point meshes in the unit cell. The energy obtained with a 13$\times$13$\times$12 k point mesh in the unit cell was taken as the reference energy $E_0$. The star symbol indicates the energy difference calculated using a 3$\times$3$\times$3 supercell with a 1$\times$1$\times$1 k point mesh.}
\end{figure}

The phonon dispersions were calculated using the Phonopy code\cite{togo2023first,togo2023implementation} based on the finite-displacement method with a supercell approach and the LO-TO splitting was included. In order to obtain convergent phonon properties in the calculations of harmonic interatomic force constants, the lattice parameter of supercell is greater than 9 {\AA}. So we made 3$\times$3$\times$3 supercell and containing 135 atoms of tetragonal PbTiO$_3$. A 20$\times$20$\times$20 q-mesh is used in the Grüneisen coefficient calculations. The Brillouin zone integration is sampled by 3$\times$3$\times$3 Monkhorst-Pack k points, other parameters are the same as those in structural optimization.

For the temperature-dependent amplitude projection analysis, AIMD simulations were performed in the NVT ensemble using a 3$\times$3$\times$3 supercell. Based on the convergence tests in Fig. \ref{Ecut_Kpot}(a) and (b), a cutoff energy of 400 eV and a 1$\times$1$\times$1 k-point ($\Gamma$ point only) were employed. Simulations were conducted for 20 ps with 10000 steps at temperatures from 100 to 700 K with a 100 K interval. The configurations between the 1000th and 10000th steps were extracted and transformed into Cartesian coordinates for amplitude projection analysis, with the sampling intervals $m$ set to 20.

\subsection{Temperature-dependent phonon dispersion}
The high-order force constants of tetragonal PbTiO$_3$ were calculated via the ALAMODE software package~\cite{tadano2014anharmonic,tadano2015self,oba2019first}. The $ab$ $initio$ molecular dynamics (AIMD) was used to calculate the vibrational configurations of each atom in tetragonal PbTiO$_3$. Based on the cutoff energy convergence test shown in Fig. \ref{Ecut_Kpot}(a), a cutoff energy of 400 eV was used to sample the vibrational configurations. The k-point sampling was chosen as 2$\times$2$\times$2, since precise k-points were not required for sampling the vibrational configurations. For the extraction of high-order force constants, detailed tests were performed to estimate the supercell size and the number of random displaced configurations. The corresponding results are presented in Sec. \ref{Precision}. The convergence tests suggest that the effects of the number of random displaced configurations on the temperature-dependent phonon dispersion (TDPD) of tetragonal PbTiO$_3$ are negligible. In contrast, the supercell size has a significant impact on the TDPD. According to this result, a 3$\times$3$\times$3 supercell and 180 random displaced configurations were selected to generate the displacement-force dataset. The magnitude of the random atom displacements was set to 0.04 \AA. The fourth order interatomic force constants (IFCs) were obtained by fitting the force-displacement datasets. The IFCs were truncated to up-to-three-body terms.

For harmonic force constants, the 13 displaced configurations were generated with the ALAMODE package. Note that the displacement magnitude of 0.002 \AA was adopted to void the appearance of imaginary phonon frequencies in PbTiO$_3$. Other settings are the same as the ab initio calculations in Sec. \ref{ab initio}.

We extracted TDPD with self-consistent phonon (SCPH) method~\cite{tadano2015self}. The SCPH calculations were carried out with an 4$\times$4$\times$4 reciprocal space k-mesh, using a 4$\times$4$\times$4 interpolating mesh. The bubble self-energy correction was also included to account for the strong anharmonicity of PbTiO$_3$. Considering the LO-TO splitting inducing abnormal imaginary frequencies of phonon dispersion at finite temperature ~\cite{eklund2024pyroelectricity}, we ignore of the LO-TO splitting in calculations of TDPD. 

\subsection{\label{Precision} Testing of temperature-dependent phonon dispersion}
PbTiO$_3$ is a classical ferroelectric material with strong long-range Coulomb interactions arising from its spontaneous polarization. Consequently, reliable calculations of third and fourth order interatomic force constants (IFCs) require sufficiently large supercells and large cutoff radii to accurately capture long-range anharmonic interactions. To generate the displacement–force datasets, equilibrium configurations of tetragonal PbTiO$_3$ at 300 K were first obtained from NVT ensemble AIMD simulations, followed by the generation of randomly displaced configurations through sampling atomic displacements from the AIMD trajectories. Since the number of sampled configurations directly determines the accuracy of the fitted higher order IFCs, the convergence of the IFCs with respect to the supercell size and the number of configurations was systematically investigated. 

\begin{figure}[H]
    \centering
    \includegraphics[width=0.7\textwidth]{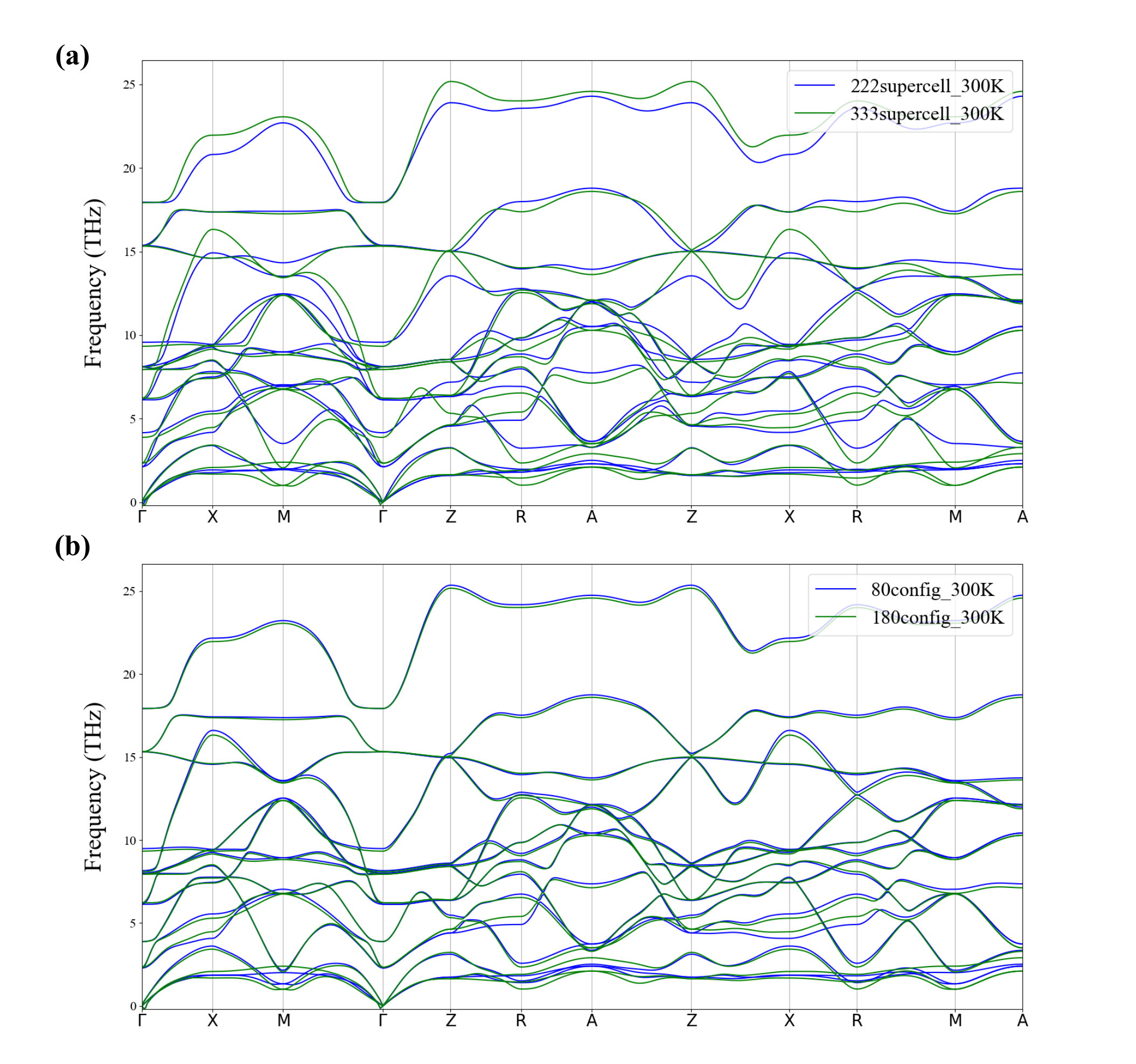}
    \caption{\label{FigS_ComparePhon} (a): Comparison of the temperature-depentent phonon dispersions calculated using 2$\times$2$\times$2 and 3$\times$3$\times$3 supercell. (b): Comparison of the temperature-depentent phonon dispersions obtained using 80 and 180 randomly displaced configurations.}
\end{figure}

To examine the convergence with respect to the supercell size, temperature-dependent phonon dispersions (TDPDs) were calculated using 2$\times$2$\times$2 and 3$\times$3$\times$3 supercells. As shown in Fig. \ref{FigS_ComparePhon}, the phonon dispersions at 300 K exhibit a dependence on the supercell size, with noticeable differences between the results obtained from the two supercell sizes. In particular, we focus on the low frequency phonon mode at the M point. The 3$\times$3$\times$3 supercell yields a lower frequency in M point phonon mode compared with the 2$\times$2$\times$2 supercell, and it is more consistent with experimental observations, where the M point mode exhibits softening behavior near the Curie temperature. Considering the balance between computational efficiency and the accurate description of long-range anharmonic interactions, the 3$\times$3$\times$3 supercell was selected for all subsequent TDPD calculations.

\begin{figure}[H]
    \centering
    \includegraphics[width=0.7\textwidth]{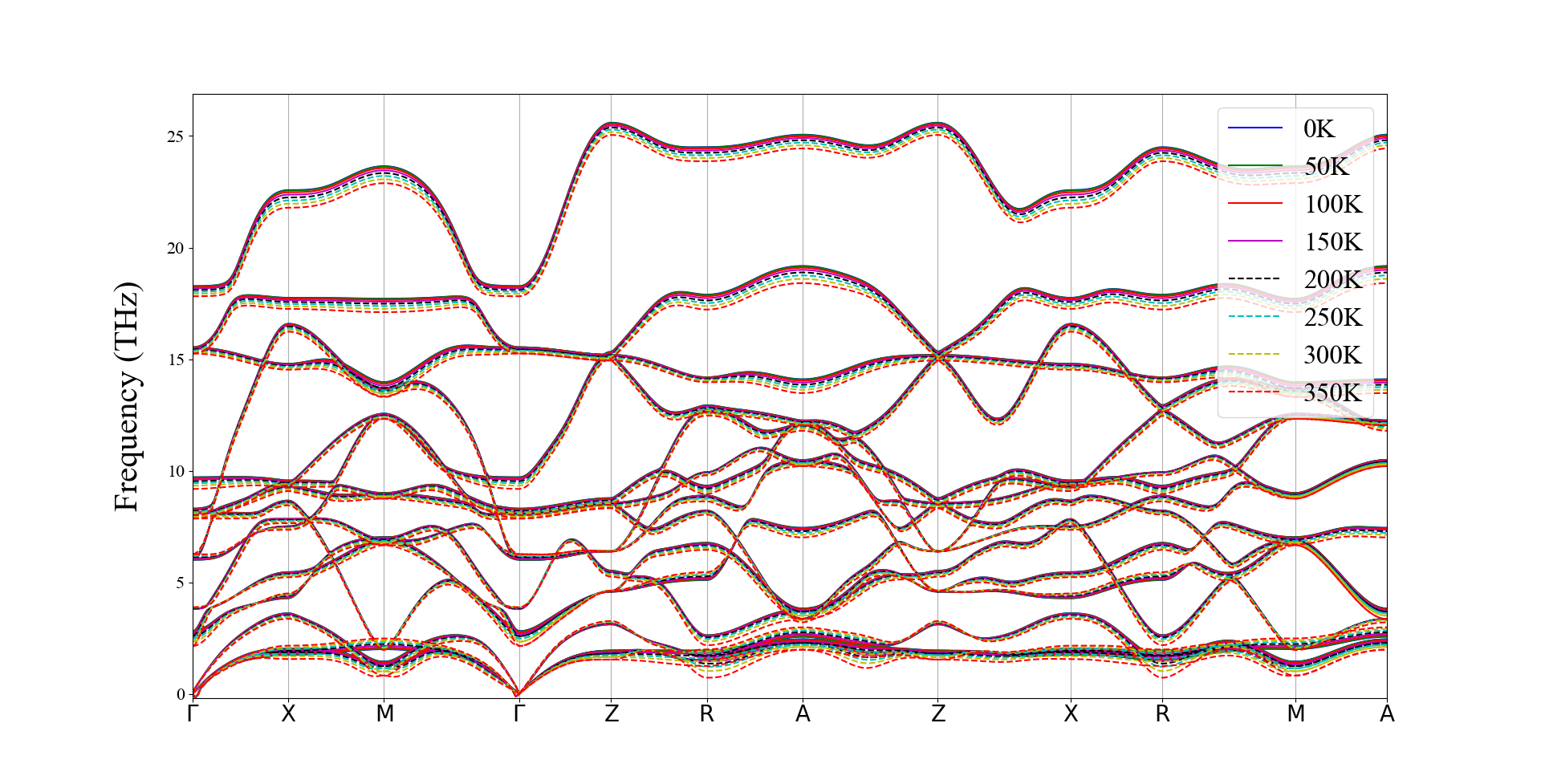}
    \caption{\label{FigS_phonon_allTemp} Temperature-depentent phonon dispersions calculated at 0-350 K (temperature interval of 50 K).}
\end{figure}

To examine the convergence with respect to the number of sampled configurations, 80 and 180 randomly displaced configurations were generated from the 3$\times$3$\times$3 supercell, and the resulting TDPDs were compared. As shown in Fig. \ref{FigS_ComparePhon}, only minor differences are observed between the phonon dispersions calculated using 80 and 180 random displaced configurations, indicating that are relatively insensitive to the number of sampled configurations within this range. Previous studies using ALAMODE have suggested that more than 100 randomly displaced configurations are generally required for reliable determination of cubic and quartic IFCs~\cite{tadano2014anharmonic,tadano2015self,oba2019first}. Therefore, 180 random displaced configurations were adopted for all subsequent calculations of third and forth order IFCs. It should be noted that the TDPDs calculated using 180 configurations show convergence issues above 350 K, whereas the phonon dispersions obtained using 80 configurations remain stable over the temperature range of 0–750 K. The deviation above 350 K can be attributed to the increasing importance of higher order anharmonic effects that are not fully captured by the fitted IFCs at elevated temperatures. Overall, the calculated TDPDs remain dynamically stable within the temperature range of 0–350 K. The TDPDs in this range, with a temperature interval of 50 K, are shown in Fig. \ref{FigS_phonon_allTemp}.

\section{\label{result} Results}
\subsection{Quasi harmonic approximation calculation of NTE}
\begin{figure}[H]
    \centering
    \includegraphics[width=\textwidth]{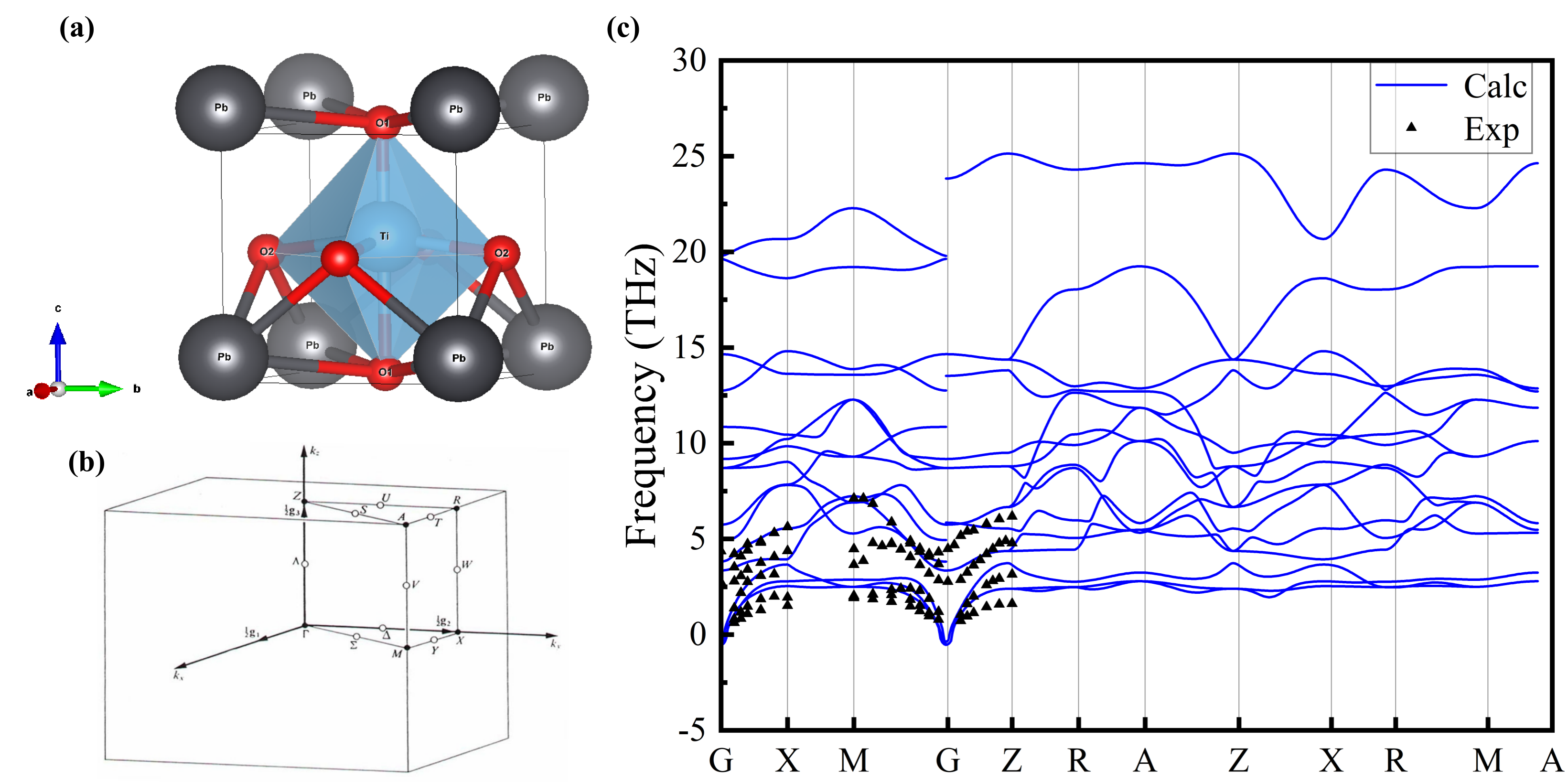}
    \caption{\label{Structure_Phonon} (a):The crystal structure of tetragonal PbTiO$_3$. (b): The First Brillouin zone with high symmetry point. (c):The phonon dispersion of tetragonal PbTiO$_3$ and compare the calculated results with the experiment~\cite{tomeno2006lattice}.}
\end{figure}

The crystal structure of tetragonal PbTiO$_3$ is shown in Fig. \ref{Structure_Phonon}(a), where Pb and Ti occupy Wyckoff positions 1a (0,0, z) and 1b (1/2,1/2, z), respectively. Among the three oxygen atoms, O1 occupies 1b (1/2,1/2, z), and O2 occupies 2c (0,1/2, z). During the structural optimization process, we test two exchange related functionals, LDA and PBE. The results are shown in supplement Table S1. It can be seen that the structure optimized by LDA functional is closer to the experimental structure, which is consistent with the results of Jiao~\cite{jiao2016first}. Therefore, LDA functional optimization of the tetragonal crystal structure will be adopted for subsequent calculations.

Then we calculate phonon dispersion by using the phonopy software package~\cite{phonopy-phono3py-JPCM,phonopy-phono3py-JPSJ}, PbTiO$_3$ was expanded with a supercell size of 3×3×3. The phonon dispersion results after considering LO-TO splitting are shown in the Fig. \ref{Structure_Phonon}(c). The coordinates of its high symmetry point are located in supplement Table S2. It can be seen that the calculated tetragonal PbTiO$_3$ exhibits negligible imaginary frequencies at the Gamma point, the negligible imaginary is from calculation error. And no imaginary frequencies in other high symmetry points. Indicating that the structure of tetragonal PbTiO$_3$ is stable at 0 K.

\begin{figure*}[ht]
    \centering
    \includegraphics[width=0.8\textwidth]{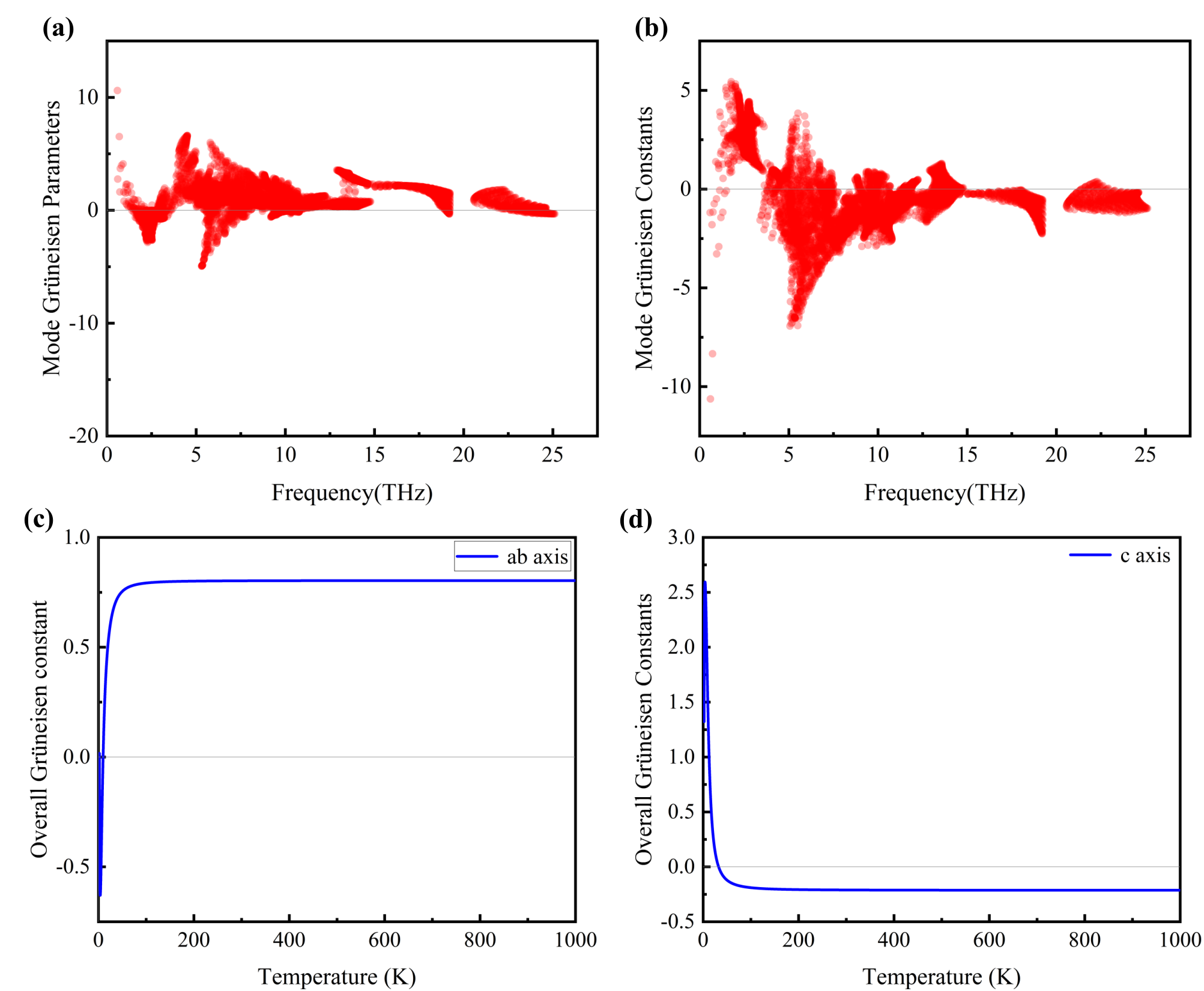}
    \caption{\label{CTE} (a): Mode Grüneisen coefficient of the $a$, $b$ axis in the Brillouin zone near the $\Gamma$ point (colors in the figure have no physical meaning); (b): Mode Grüneisen coefficient of the c axis in the Brillouin zone near the $\Gamma$ point; (c): Overall Grüneisen constants allong $a$ and $b$ axis; (d): Overall Grüneisen constants allong $c$ axis.}
\end{figure*}

Fig. \ref{CTE}(a) and (b) show the mode grüneisen coefficients in the Brillouin zone near the Gamma point.
The mode grüneisen coefficient of the $a$ and $b$ axis shows negative values near frequencies of 5 THz, corresponding to the negative mode grüneisen coefficients at points M and A marked in Fig. 1(a) of the main text. The c axis exhibits negative mode grüneisen coefficients in the frequency of 5 THz, and higher frequencies. Among them, the mode grüneisen coefficients in the frequency range of 5THz and above account for the majority of the negative mode grüneisen coefficients in the entire result. According to Eq. \ref{Eq_CTE}, it can be seen that the magnitude of the system's axial NTE is proportional to the negative overall uniaxial Grüneisen coefficients, and according to Eq. \ref{Eq_OverallGruneisen}, the uniaxial Grüneisen coefficients reflect the proportion of negative mode grüneisen coefficients in the entire Brillouin zone. Therefore, it can be concluded that the c axis NTE of tetragonal PbTiO$_3$ is dominated by the optical branch, and the low frequency optical branch has a significant impact on the c axis NTE. This result is consistent with Wang et al.\cite{wang2014first} are consistent: the optical branch promotes volume contraction.

After that, we calculated the flexibility coefficient of PbTiO$_3$, and the results are shown in Tab. \ref{Tab_felxiblity}. The overall (uniaxial) Grüneisen coefficients are shown in Fig. \ref{CTE}(c) and (d). The uniaxial Grüneisen coefficients along the a and b axes are positive above 100 K, whereas those along the c axis are negative. 

\subsection{A point vibrational mode of TiO$_6$ octahedron and amplitude projection}



\begin{figure}[H]
    \centering
    \includegraphics[width=0.65\textwidth]{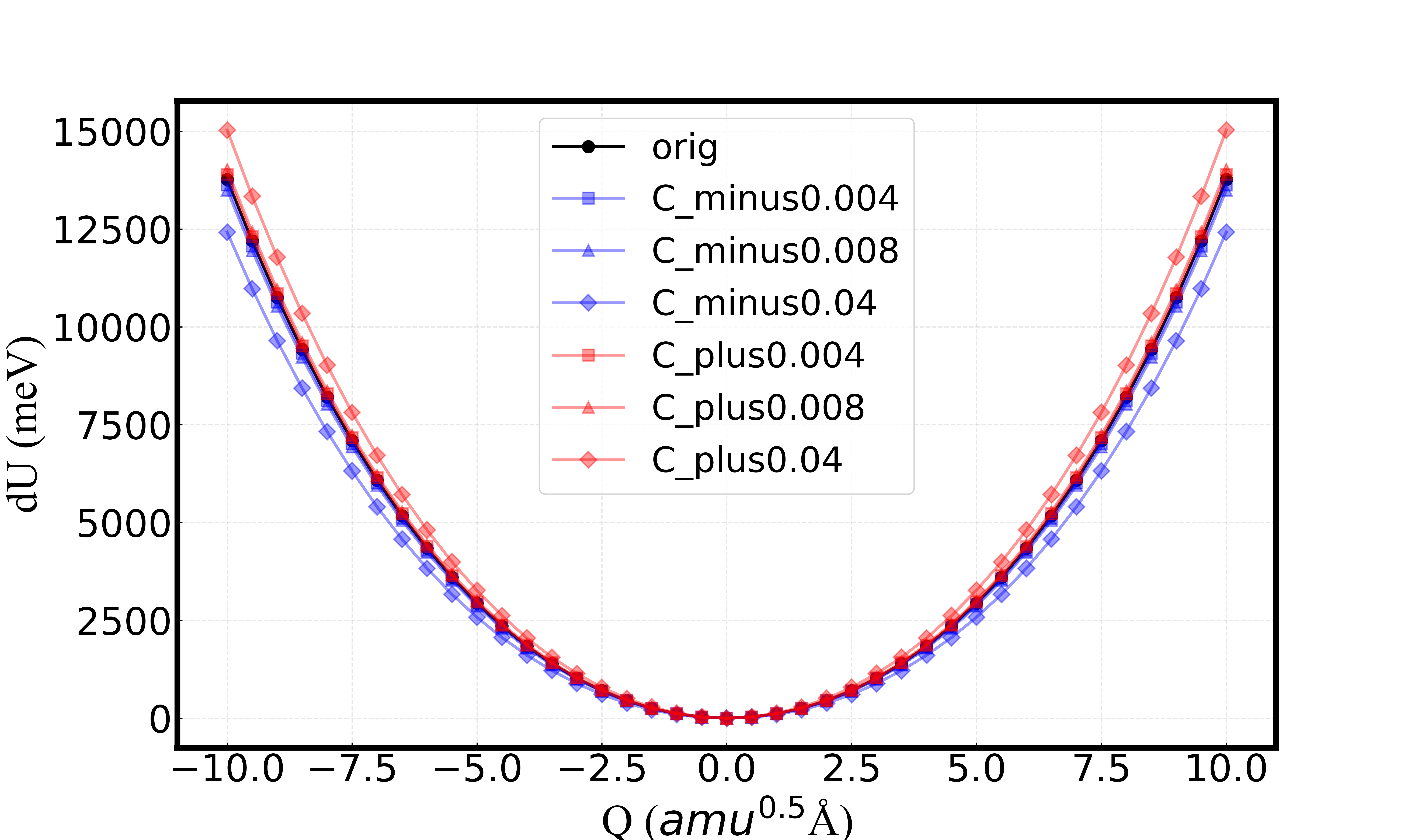}
    \caption{\label{FigS_Q_DE} Potential energy surface of TiO$_6$ octahedral rotational mode at A point after c axis compressed and stretched.}
\end{figure}

\begin{figure}[H]
    \centering
    \includegraphics[width=0.8\textwidth]{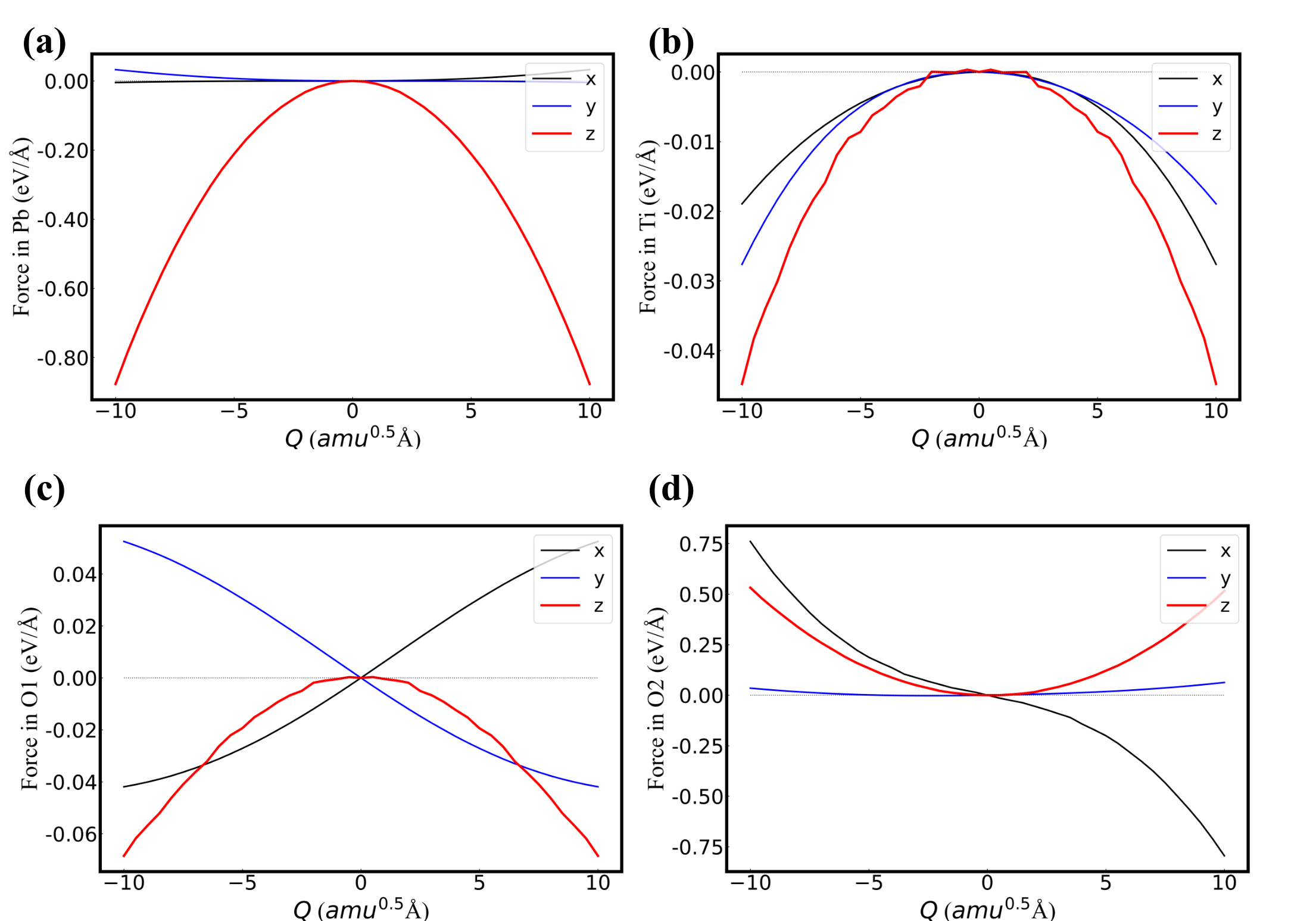}
    \caption{\label{FigS_Force_Apoint} (a)-(d): Influence of the A point TiO$_6$ octahedral rotational mode on atomic forces. The force acting on Pb, Ti, O1 and O2 atoms as a function of the amplitude of the A point TiO$_6$ octahedral rotational mode. The atoms is signed in Fig. 3(g) of the main text.}
\end{figure}

\begin{figure}[H]
    \centering
    \includegraphics[width=0.8\textwidth]{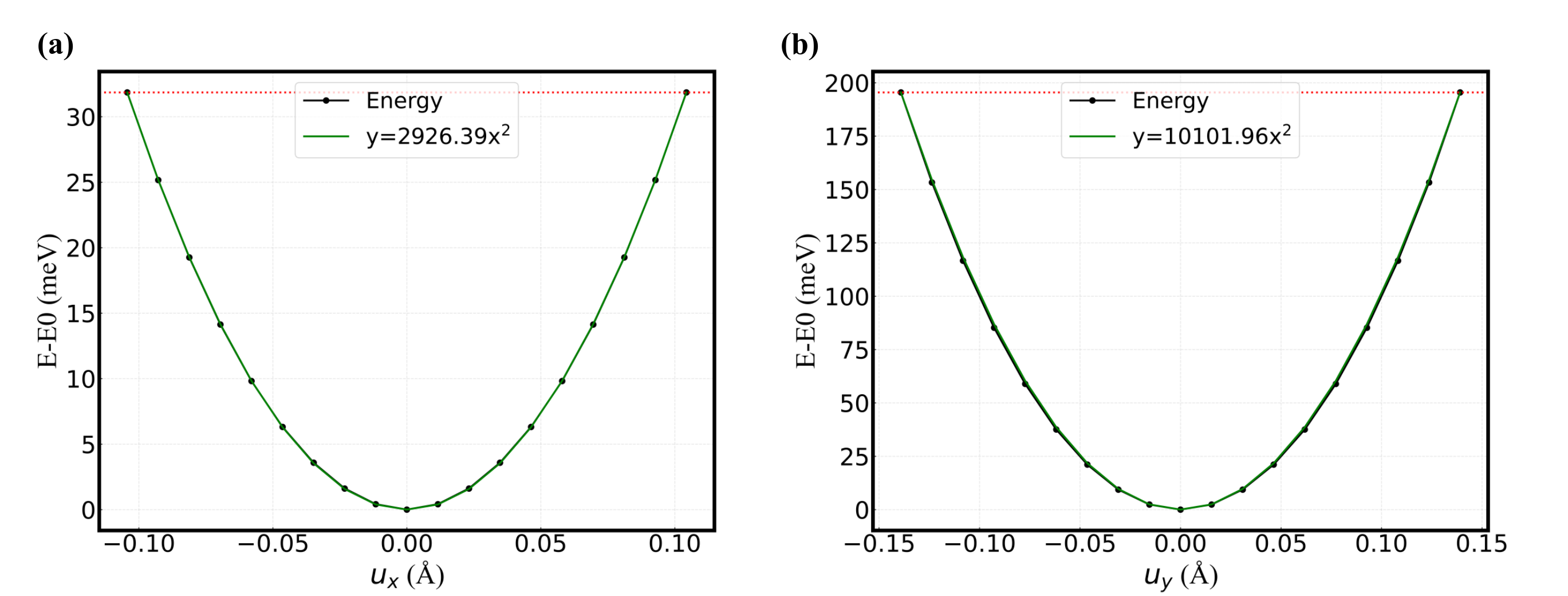}
    \caption{\label{FigS_vibration_xy} (a): Energy difference of the O2 atoms displaced along the $x$ direction. The green line in the figure representing the reference parabola. (b): Energy difference of the O2 atoms displaced along the $y$ direction.}
\end{figure}

\newpage
\bibliographystyle{model1-num-names}

\bibliography{apssamp}